\documentclass[12pt]{article}

\usepackage{scicite}

\usepackage{times}
\usepackage{url}

\usepackage{amsmath}
\usepackage{amsfonts}
\usepackage{graphicx}
\usepackage{comment}
\usepackage{booktabs}
\usepackage{threeparttable}
\usepackage{multirow}
\usepackage{mdwtab}

\usepackage[normalem]{ulem}
\newcommand\redout{\bgroup\markoverwith
{\textcolor{red}{\rule[0.5ex]{2pt}{1pt}}}\ULon}

\usepackage{xcolor}

\usepackage{caption}
\newenvironment{sciabstract}{%
\begin{quote} \bf}
{\end{quote}}

\title{Urban Dynamics Through the Lens of Human Mobility}

\author
{Yanyan Xu$^{1,2,3}$, Luis E. Olmos$^{2}$, David Mateo$^4$, Alberto Hernando$^4$, \\ Xiaokang Yang$^{1}$, Marta C. Gonz\'{a}lez$^{2,3,5\ast}$\\
\\
\normalsize{$^{1}$MoE Key Laboratory of Artificial Intelligence, AI Institute,}\\
\normalsize{Shanghai Jiao Tong University, Shanghai 200240, China}\\
\normalsize{$^{2}$Department of City and Regional Planning, University of California,}\\
\normalsize{Berkeley, CA 94720, USA}\\
\normalsize{$^{3}$Energy Technologies Area, Lawrence Berkeley National Laboratory,}\\
\normalsize{Berkeley, CA 94720, USA}\\
\normalsize{$^{4}$Kido Dynamics SA, 1004 Lausanne, Switzerland}\\
\normalsize{$^{5}$Department of Civil and Environmental Engineering, University of California,}\\
\normalsize{Berkeley, CA 94720, USA}\\
\\
\normalsize{$^\ast$To whom correspondence should be addressed. E-mail: martag@berkeley.edu.}
}

\date{}

\begin{document} 
\baselineskip24pt
\maketitle 

\begin{sciabstract}
The urban spatial structure represents the distribution of public and private spaces in cities and how people move within them. While it usually evolves slowly, it can change fast during large-scale emergency events, as well as due to urban renewal in rapidly developing countries. This work presents an approach to delineate such urban dynamics in quasi-real-time through a human mobility metric, the mobility centrality index $\Delta KS$. As a case study, we tracked the urban dynamics of eleven Spanish cities during the COVID-19 pandemic. Results revealed that their structures became more monocentric during the lockdown in the first wave, but kept their regular spatial structures during the second wave. To provide a more comprehensive understanding of mobility from home, we also introduce a dimensionless metric, $KS_{HBT}$, which measures the extent of home-based travel and provides statistical insights into the transmission of COVID-19. By utilizing individual mobility data, our metrics enable the detection of changes in the urban spatial structure.
\end{sciabstract}


Urbanization is arguably the most important change in recent history, which greatly transformed how people live, work, and travel~\cite{bettencourt2013origins,batty2013theory}. Approximately 55\% of the world population lives in urban areas and this proportion has reached 80\% in high-income countries~\cite{un2018world}. However, the rapid urban expansion imperils socio-economic and environmental well-being with consequences livability~\cite{makse1995modelling,batty2008size,barthelemy2016structure,keuschnigg2019urban,xu2020deconstructing}. For example, economic growth and the production of new inventions~\cite{bertaud2004spatial,bettencourt2007growth} scale with city size, but urban expansion can also introduces negative effects such as the exacerbation of regional air pollution and urban heat island effects~\cite{ewing2008impact,lamsal2013scaling,ribeiro2019effects,li2019urban}.
During the process of urbanization, the urban structure affects the environmental and energy costs of settlements~\cite{anderson1996urban,bento2005effects,tsekeris2013city,kaza2020urban}. Specifically, it confines the development of a city into a given space, including the planning of transportation, and the expansion of the labor market. In light of this, urban designers often seek to regulate the urban form towards long-term sustainable targets.

Yet, the evolution of urban spatial structure, namely urban dynamics, is only slowly monitored, because the complex interaction between population, resources, and urban regulations makes the collection of the required information a laborious work. In urban planning, urban form is commonly defined using two complementary components: (i) the spatial distribution of the population~\cite{clark1951urban,newling1969spatial,bertaud2003spatial}, and (ii) the spatial distribution of functional facilities or working places~\cite{pereira2013urban}. In the absence of mobility data, researchers generally utilize census data about employment or commuting flows as a proxy for trip attractors~\cite{sohn2005commuting,nelson2016economic}.
However, national or state-level travel surveys are usually conducted once every few years, which is insufficient to provide up-to-date information.
For example, census data fails to take into account the frequent occurrences of large-scale emergency events, such as pandemics, natural disasters, or socio-economic crises. These events might lead to relocations and changes in people's mobility patterns~\cite{gray2012natural,lu2012predictability,acosta2020quantifying}. Disasters are generally accompanied by economic disruptions~\cite{hsiang2017estimating}, which further trigger changes in socio-economic activities. All these changes in housing and socio-economic interactions would reshape the urban structure. Therefore, it is necessary to identify the urban dynamics in a quasi-real-time manner, to facilitate timely policy responses and more effective planning~\cite{smigiel2019urban,xu2020deconstructing,kasmalkar2020floods}.

In the past decade, massive mobility data collected from dwellers have been a powerful tool for urban planning, as data-driven schemes unravel the dynamic interaction between humans and their complex urban systems~\cite{calabrese2014urban,barbosa2018human,olmos2018macroscopic,xu2018planning,yabe2022toward}. The sharing of individual trace data has raised privacy concerns~\cite{rocher2019estimating,grantz2020use,oliver2020mobile}. For example, by studying massive mobile phone data, researchers found that the uniqueness of mobility traces can reach up to 95\%, when users are localized by the operator's antennas~\cite{de2013unique}.  
Here we argue, however, that there are still important insights to gain from individuals' mobility data, particularly in generating metrics that can quantify instant urban structures in the face of catastrophic events~\cite{anderson2018dangers,yabe2022toward}.

We study multiple cities and explore the potential of defining urban spatial structure with human mobility. We observe that the distribution of housing with respect to the CBD is a decisive factor in determining the scale of mobility. For example, we examine the distributions of the home-centered radius of gyration $Rg$~\cite{gonzalez2008understanding} extracted from disaggregated mobile phone traces in Boston, Los Angeles (LA), and Bogot{\'a}. We observe that the wealthiest group in Bogot\'a tend to reside in a region near the CBD, with shorter $Rg$, whereas, in cities such as Boston and LA, wealthier groups reside far away from the CBD and travel longer distances. Economic segregation and the relative position of each group with respect to the CBD explain the distance traveled within the urban borders. Yet, when taking into account trips outside the city during the holidays, $Rg$ is generally larger for individuals with higher incomes.

Motivated by this finding, we propose a metric of human mobility as a function of the distance of the travelers' group from the CBD. Via studying twenty-one cities from America, Asia, and Europe (Figs.~S1 and S2). We calculate the $Rg$ distributions in multiple circular rings centered at the CBD and propose a mobility centrality index, $\Delta KS$. The $\Delta KS$ quantifies how strongly people's radius of gyration changes depending on the distance from the city center. The statistical divergence between $Rg$ values of different groups of people is measured with the Kolmogorov–Smirnov (KS) statistic. We analyze twenty-one cities and observe that our scale-independent metric is capable to assess a wide spectrum of urban structures ranging from monocentric to polycentric forms. More specifically, a large $\Delta KS$ indicates a monocentric urban structure, suggesting that people frequently travel to CBDs even if they live in the periphery. While a more homogeneous distribution of $Rg$ indicates that a city is more polycentric, where various commercial centers have emerged. Our observed values of $\Delta KS$ usually range from 0 to 1. Specifically, $\Delta KS$ approaches $0$ if the trips are uniformly distributed across the city.

We next explore the change in people's mobility behavior during a large-scale emergency, the COVID-19 pandemic, in Spain. At the onset of the COVID-19 outbreak over the globe in early 2020, to assist policymakers and researchers to propose reasonable prevention and control strategies, several location intelligence companies released their mobility data to the academic community~\cite{bonaccorsi2020economic,chang2020mobility,oliver2020mobile,grantz2020use,engebretsen2020time,nouvellet2021reduction,alessandretti2022human}. There are several works that have addressed the impact of mobility in the spread of epidemics. For example, Linka et al. attempted to predict the dynamic of COVID-19 at a national scale with local car traffic and global air mobility~\cite{linka2021global}. Schlosser et al. investigated the impact of the COVID-19 epidemic on mobility flow networks across cities in Germany and discovered that the connection between regions was weakened due to a significant reduction in long-distance trips~\cite{schlosser2020covid}. In addition, the mobility needs of the population can serve as an early warning sign of COVID-19 outbreaks when combined with other digital traces.~\cite{kogan2021early,stolerman2023using}. In this context, we propose urban dynamics metrics that do not depend on the urban scale, facilitating the comparative study in multiple cities.

Using disaggregated mobile phone data from 17 million anonymized users in Spain in 2019 and 2020, we couple the time-varying $Rg$ with their annual income and unveil the inequitable impacts of COVID-19 pandemic by looking at the mobility behavior of different income groups. We observe that the population with lower income took longer trips during the first and second waves in 2020, likely reflecting they could not afford to stay at home. In contrast, in the holiday seasons of April and August 2019 higher income groups have longer trips.

Further, we select eleven cities in Spain to explore their urban dynamics during the eight months after the outbreak of COVID-19 in 2020, as well as the function of mobility behavior and urban dynamics in the epidemic spreading. Based on the daily reported figures of infections, we divide the study period into three phases, the first wave (before May 1), the quiet phase (from May 1 to July 1), and the second wave (after July 1). We find that all of the studied cities became more monocentric during the first wave of COVID-19, indicating that that trips with destinations outside the CBDs decreased.
After the first wave, most cities' urban structures recovered to their usual levels in June 2020, and kept stable during the second wave. 
As an exception, Alicante stayed more polycentric after June, showing that trips to its CBD had not recovered by the end of 2020. 
In this context, we propose a home-based travel indicator $KS_{HBT}$ to measure the extent of mobility restriction during an emergency. $KS_{HBT}$ measures the statistical distance between the actual $Rg$ and the $Rg$ if all population stayed at or near home. 
Via modeling the effective reproduction number $R_t$, we quantify the impacts of the time-varying mobility behavior and urban structure on the spread of COVID-19. Our results confirm that the mobility metrics $Rg$ and $KS_{HBT}$ are the most important factors to model $R_t$, followed by the mobility centrality index $\Delta KS$. Especially during the first wave, mobility variables present a stronger positive impact on $R_t$. 
All the variables and notations introduced in this work are summarized in Note~S1.

\section*{Urban Space through the Lens of Racial and Economic Status}
Spatial segregation curtails the opportunities for people to access their needed resources, such as education, jobs, and public facilities. This in turn enlarges the income gaps between different groups and even affects the health outcomes among different levels~\cite{bor2017population}. By analyzing mobility data, we find that cities display socioeconomic segregation patterns, independent of their specific urban structure. Fig.~1A presents racial and income groups in Boston and LA in the United States, as well as the map of Bogot\'a, Colombia, marked by the socio-economic strata (SES). Details of the data sets are described in \textit{Datasets description} in Materials and Methods and Note~S2. As expected, residents are segregated by race and income in the subject cities. 
Similar to other cities in the US~\cite{brueckner1999why}, both Boston and LA exhibit a pattern where higher-income groups reside on the periphery (light colors) while lower-income groups reside closer to the city center (dark colors). In contrast, Bogot\'{a} displays a different pattern, with higher-income groups (SES 5, 6) located in the northern part of the city and lower-income groups (SES 1, 2) on the periphery. The relationship between income levels and distance from the CBD for the three cities is shown in Fig.~S3.

We next investigate the mobility patterns of mobile phone users from diverse racial and economic backgrounds. 
To reconstruct individual users' trip chains, we use the TimeGeo modeling framework to complete the footprints provided by call detailed records (CDRs)~\cite{jiang2016timegeo}, see Note~S3. Simulation results of TimeGeo in some of the cities have been validated with travel survey data~\cite{jiang2016timegeo,xu2018planning,de2020socio}.
For the cities with high-frequency extended detail records (XDRs), we identify the significant stay locations with the raw trajectory data as XDRs are more frequently collected and keep almost all locations visited by the users. Besides, we use the travel survey data in Atlanta and Rio de Janeiro (see \textit{Datasets description} in Materials and Methods).
By defining an information theory segregation index $H$, interpreted as the population-weighted difference between the diversity of each spatial unit and the diversity of the whole city, we notice that the spatial segregation is significantly lower during the daytime than during the night (Note~S4 and Fig.~S4A). This highlights how people's residential location plays a dominant role in the spatial segregation observed in these cities. Interestingly, when we inspect the non-work activities, results display that the majority of the visitors in Boston and LA are from the same racial group as their destinations, see Figs.~S4B-D. Similar patterns are observed for the SES in Bogot\'a. These observations establish that the socio-economic segregation during leisure time captures homophily in mobility choices.

Urban segregation naturally leads to divergence in mobility behavior~\cite{louf2014congestion,florez2017measuring,wang2018urban,moro2021mobility}. This divergence comes from not only people's residential location but also from the low availability of jobs and facilities in peripheral regions~\cite{simini2012universal,carra2016modelling}. Figs.~1B-D presents the distribution of the radius of gyration, $Rg$, of different socio-demographic groups in Boston, LA, and Bogot\'a. $Rg$ captures the spatial coverage of each user's daily mobility, centering at her home location~\cite{gonzalez2008understanding}. Considering a user's mobility behavior during a certain period as a sequence of visited locations, then $Rg$ is calculated as
\begin{equation}
    Rg = \sqrt{\sum_{i=1}^n{\frac{1}{n}(\boldsymbol{l}_i-\boldsymbol{l}_h)^2}}
\end{equation}
where $n$ is the length of the sequence, $\boldsymbol{l}_i$ and $\boldsymbol{l}_h$ are the geographical coordinates of the $i$th visited location and the home location, respectively. We then average individual's daily $Rg$ values over the various data availability periods ranging from one month to six months. The relation between people's $Rg$ and their household income levels for the aforementioned three cities is presented in Fig.~S5. In conjunction with Figs.~1B-D, we observe that the inhabitants with the highest income tend to travel over longer distances in Boston and LA, whereas inhabitants with the lowest income travel over longer distances in Bogot\'a. Therefore, there is no coincidental relation between the distribution of $Rg$ and the socio-economic characteristics in diverse cities. The reason is that each group's mobility behavior is mainly determined by spatial segregation, which is driven by the socio-economic interactions and some even more complex dynamics that brought up the locations of various groups in reference to the CBDs.

During the COVID-19 pandemic, governments around the world imposed travel restriction policies to a certain extent, causing economic activities to stagnate and the unemployment rate to increase~\cite{bonaccorsi2020economic,verschuur2021observed}. Different socio-economic groups reacted to COVID-19 and the consequent non-pharmaceutical interventions (NPIs) in various degrees~\cite{iio2021covid}. With the data of cities across different countries, researchers observed a greater decrease in the mobility of people with higher socio-economic status~\cite{hou2021intracounty,brough2021understanding,kar2022essential}. 
We confirm this finding at a national scale, as shown in Fig.~1E. We depict the relation between the individuals' annual income and the daily $Rg$ of the population in Spain during 2019 and 2020, covering the period before and during the COVID-19 pandemic. We identify that long-distance trips taken by high-income residents in April 2019 disappeared in the plot of 2020. In addition, comparing the months before and after March 1, 2020, the travel distance of high-income groups drops more significantly than that of the low-income groups during the early-stage of COVID-19, as well as in the second peak season from October to December in 2020. Between June and October 2020, travel restriction policies became relaxed, and trip distances returned to their usual values. Taken together, high-income groups usually take more long-distance trips during tourist seasons than low-income groups do, while this situation got reversed during the peak seasons of COVID-19 pandemic. This might be because there is a higher fraction of essential trips in the low-income population with lower numbers working from home.

\section*{Proposed Mobility Metrics}

In daily lives, humans take trips to work, access resources or attend social activities in urban complex systems. Usually, these trips are centered around residential locations as illustrated in Fig.~2A. Meanwhile, cities have their organization schemes, which can be abstracted as a spectrum ranging from monocentric to polycentric forms. In monocentric cities, facilities are concentrated to the city centers, while in polycentric cities, facilities are distributed much more dispersedly~\cite{xu2020deconstructing}. 

Fig.~2B schematically depicts the typical mobility behavior of people residing in monocentric and polycentric cities. Typically, in a monocentric city, people with homes closer to the CBD have more accessibility to various resources, thereby they travel over shorter distance than those living in the peripheral area. In a polycentric city, people can access their needed resources in the sub-centers as well. We confirm this expectation empirically with mobility data from multiple cities. Specifically, the spatial distribution of the average $Rg$, at the census tract level, discloses that people's place of residence has a significant impact on the spatial scale of their daily activity. Fig.~S6 depicts the spatial distribution of $Rg$ values in Boston, LA and Bogot\'a. We notice that, unlike other regions, the southern region of LA keeps a small $Rg$, even if it is far away from the CBD. That is in conformity to LA's polycentric urban structure. As illustrated in Fig.~2B, we can expect that monocentric cities have a more divergent distribution of $Rg$, while the distributions of $Rg$ are expected to be more uniform for polycentric cities, due to the dispersed distribution of job opportunities and resources.

Inspired by Bertaud and Malpezzi's work on urban form identification with population distribution at varying distances to the city center~\cite{bertaud2003spatial}, we divide a city's population into multiple groups based on the distances of their residential locations to the CBD. As illustrated in Fig.~2C, we delimit a sequence of concentric rings centered at CBD with a constant width. The collection of $Rg$ of the population in the most inner CBD circle and the $i$th ring with internal radius $r_i$ are denoted as $\left\langle Rg(r_0)\right\rangle$ and $\left\langle Rg(r_i)\right\rangle$, respectively. 
Next, we attempt to fit the relationship between the average $Rg$ and the radius of the ring using a linear function. Our analysis reveals that while there is an obvious linear relationship in some cities, this is not the case in others, as shown in Fig.~S7. Fig.~S8 summarizes the average and median values of $Rg$ in concentric rings at different distances from the CBD for the 21 cities. The results indicate that, regardless of how the cities are organized, individuals residing farther from the CBD tend to take longer trips than those living near the CBD in most cities. From this perspective, the average or the median $Rg$ as a function of the distance to the CBD can not provide enough information to define urban spatial structure.

Here we quantify the divergence between $\left\langle Rg(r_0)\right\rangle$ and $\left\langle Rg(r_i)\right\rangle$ with the Kolmogorov–Smirnov (KS) statistic, as shown in Fig.~2D. That is,
\begin{equation}
    KS(r_i|r_0) = \mathbf{KS} \left( \left\langle Rg(r_i)\right\rangle, \left\langle Rg(r_0)\right\rangle \right) = \sup_{Rg} \left | F_{\left\langle Rg(r_i)\right\rangle} - F_{\left\langle Rg(r_0)\right\rangle} \right |
\end{equation}
where $\mathbf{KS}(\cdot)$ denotes the KS distance that quantifies the statistical divergence between two groups of data with their cumulative distributions. $F_{\left\langle Rg(r_0)\right\rangle}$ and $F_{\left\langle Rg(r_i)\right\rangle}$ are the cumulative distribution functions of $\left\langle Rg(r_0)\right\rangle$ and $\left\langle Rg(r_i)\right\rangle$ respectively, and $\sup$ is the Supremum function. Some other metrics to measure the statistical divergence of $Rg$ values between two groups of people are discussed in Note~S5.

Empirically, as mentioned above, the urban structure and the residential location are two major factors in determining people's mobility radii. Accordingly, we introduce the concept of mobility centrality, an index to quantify the heterogeneity on the mobility scale of the population residing in different areas. To represent cities with varying sizes, we scale each city into a circle of radius 1 with its CBD being the center. Specifically, we first define $r_{max}$ of each city as the radius of the smallest circle which covers at least $95\%$ of the total population. Then the relative radius $\hat{r}$ is defined as the ratio between the actual radius $r$ and the maximum radius of the city $r_{max}$. Next, as illustrated in Fig.~2E, we fit the relation between the radii of the rings $\left\langle \hat{r}_i \right\rangle$ and their $\left\langle KS(\hat{r}_i|r_0) \right\rangle$ with a linear function, then name the slope $\Delta KS$ as the mobility centrality index to assess the urban structure. 
It is noteworthy that $\hat{r}$ ranges from 0 to 1 and $KS(\hat{r}|r_0)$ is smaller than 1, we can expect $\Delta KS$ to usually range between $0$ and $1$ in real-world cities, independent of their spatial scales. We can also expect that $\Delta KS$ can approach $0$ if all of the required resources are uniformly distributed across the city, and it will be large if people's distance to CBD heavily impacts their mobility behavior. In this context, a larger $\Delta KS$ means that the scale of people's mobility increases faster regarding their distances to CBD, suggesting a more monocentric urban form.

As a major driving force of virus spreading, human mobility received unprecedented attention from researchers and policymakers during the early stage of the COVID-19 pandemic~\cite{chinazzi2020effect,kraemer2020effect}. Many important NPIs take mobility into account. For example, the effect of shelter-in-place order was generally measured via the travel flow or the distance of displacement~\cite{pepe2020covid,oliver2020mobile,grantz2020use}. Also, the widely used aggregated flow and average distance assess diverse aspects of human mobility while keeping individual trajectories private. However, these indicators have some drawbacks. First, disparate distributions of travel behavior may share the same flow or average distance. In addition, flow and distance depend greatly on a city's population and scale, hindering the comparison of mobility behavior across regions.
To remedy this, we propose $KS_{HBT}$, namely extent of home-based travel, as a way to measure behavior disparity of the population residing in different regions, to assess the effect of shelter-in-place behavior. As illustrated in Fig.~2F, $KS_{HBT}$ refers to the $KS$ distance between (i) the distribution of $\left\langle Rg_{actual} \right\rangle$, the observed $Rg$ of all individuals including $Rg=0$ and (ii) the distribution of $\left\langle Rg_{home} \right\rangle$, the $Rg$ if all individuals took the shelter-at-home order. For the latter, we assume that $Rg$ follows a uniform distribution between $0$ and a predefined threshold $r_{shelter}$. $r_{shelter}$ indicates the mobility radius of people following the travel restriction order and is set as 0.5~km in this work, which is nearly the median $Rg$ in all eleven Spanish cities during the lockdown (see Fig.~S16). As can be seen in Fig.~2F, $KS_{HBT}$ can be explained as the share of the population with $Rg$ over $r_{shelter}$. Therefore, a lower $KS_{HBT}$ means that more people stay at or near their homes.

\section*{Urban Dynamics via Individual Mobility Metrics}
Urban structure is currently monitored with the distributions of population or employment~\cite{anas1998urban}. Inspired by Meijers and Burger's work~\cite{meijers2010spatial}, Fig.~3A depicts the two dimensions of urban spatial structure, one dimension quantifies the population distribution, ranging from dispersion to compaction, and the other represents the centrality of functional units in space, ranging from polycentricity to monocentricity. The population distribution of sample cities is represented with 3D barplots. Population in blocks with a size around 1~km$^2$ comes from the LandScan dataset~\cite{landscan2015} and the visualization of all studied cities is presented in Fig.~S2.

We illustrate the distribution of $Rg$ in each ring with $3$~km width for four cities (Shenzhen, Wuhan, LA, and Boston) in Fig.~3B. Similar plots for other cities are presented in Fig.~S9. For Wuhan and Boston, as the distance from the CBD increases, the $Rg$ distribution shifts to a larger average value. While for Shenzhen and LA, $Rg$ displays a similar distribution agnostic of the distance from the CBDs. 
This indicates that from the perspective of mobility, Wuhan and Boston are both monocentric. The two cities both have large proportions of the trips attracted to their CBDs, though the population distribution in Wuhan is more disperse than Boston. In turn, Shenzhen and LA are both polycentric cities from the point of view of human mobility, as people's movement is dispersed among multiple destinations, although the population distributions of the two cities are different. Shenzhen's population distribution is more compact, with the majority of residents living within the CBD circle, while LA has a more uniform distribution of population across the city. Therefore, we can expect that measuring the statistical divergence of people's $Rg$ indicates polycentricity and complements well the information drawn from population distributions.

Fig.~3C presents the $KS(\hat{r}|r_0)$ of twenty-one cities. It quantifies the changes in the distributions of $Rg$ away from the CBDs. Note that as we are using $KS(\hat{r}|r_0)$ to measure the disparity of human mobility behavior, we exclude the users who didn't move during daytime ($Rg=0$).
A constant value as a function of $\hat{r}$ suggests that the daily mobility of the population residing in different regions appears to have similar $R_{g}$ distributions. We propose a mobility centrality index $\Delta KS$, as one metric to capture the change of $KS(\hat{r}|r_0)$ in space (shown in Fig.~2E). We use $\Delta KS_{typ}$ to represent the typical values of $\Delta KS$ over long-term periods of observations, varying from one to six months in our twenty-one studied cities. For the Spanish cities, we calculate $\Delta KS_{typ}$ based on the average daily $Rg$ of the population during October 2019. For other cities, $\Delta KS_{typ}$ is calculated over the various periods of observation when data is available (details in note~S2).
The results of the fit are presented in Fig.~S12 with cities sorted in descending order of $\Delta KS$. The goodness-of-fit is also presented in Table~S1. We observe that the $r^2$ is reasonable in multiple polycentric cities (LA, Shenzhen, etc.), except for SF Bay Area, which has an irregular spatial topology in comparison with the others.

Regarding population distribution, we calculate the Gini index to measure the uniformity across urban blocks with a population of over 500 people, which is a common threshold to distinguish between urban and rural regions~\cite{xu2020deconstructing}. A larger Gini value indicates that a city is more compact in population distribution, while a lower value means a city's population is more dispersed.
In Fig.~3D, we plot the Gini index vs. $\Delta KS_{typ}$ and find that $\Delta KS$ mildly increases with the Gini index of population, in line with the intuition that the more compact cities tend to be more monocentric (Fig.~S13). For example, Wuhan, where the COVID-19 pandemic began in December 2019, is the most compact in terms of Gini index. The $\Delta KS_{typ}$ of Wuhan is also the second-largest, after Santa Cruz. In comparison with Wuhan, cities like Porto, Alicante, Lisbon, and Zaragoza, have similar $\Delta KS_{typ}$ but are more dispersed in population distribution, suggesting that they may be undergoing processes of urban sprawl~\cite{martins2012urban,rubiera2016urban}. Shenzhen is a crowded city, with a large population in a relatively small area, 10.72 million population residing in 1,956 km$^2$ region. Shenzhen's population peaks (compactly) in several districts, and the uniform distribution of job opportunities and resources shape it into a polycentric form. Other polycentric regions, such San Francisco Bay Area (SFBay) and LA, are dispersed regarding their population distribution. 

Note that, in the analysis above, we set the constant width of rings as $3$~km. We tested our definition of urban structures with different ring widths, as presented given in Fig.~S14. Results show that, though $3$~km ring width is slightly better than 1 and 2 km widths in terms of r-square of the linear fit, different ring widths do not affect the conclusions of the results (Fig.~S14C-G).

\section*{Sensing Urban Dynamics in Times of Crisis}

During February 2020, cities in Spain were successively hit by the COVID-19 pandemic and the government imposed a nationwide mandatory lockdown on March 14. We use individual mobile phone data collected from February to September 2020 to dissect the changes in mobility behavior and urban dynamics during such a period in eleven major cities in Spain. Fig.~4A shows the daily infection incidence in each province per 1,000 people before September 30, 2020. The numbers of newly confirmed infections are presented in Fig.~S15. The study period can be broken down into three phases: phase I covers the first wave of COVID-19 before April 30, phase II covers the quiet period from May 1 to June 30, and phase III covers the second wave after July 1. 

With the observed individual $Rg$, we calculate the daily metrics $\overline{Rg}^d$, $\Delta KS^d$, and $KS_{HBT}^d$, per city, from February 1 to September 30, 2020, as depicted in Fig.~S17. All three mobility metrics display a weekly periodicity. To curb the behavioral fluctuation caused by weekdays, we further average each mobility metric during seven consecutive days, namely $\overline{Rg}^{7d}$, $\Delta KS^{7d}$, and $KS_{HBT}^{7d}$, as presented in Fig.~S18. We present the relative changes of the three variables in Figs.~4B-D, in relation to their values in the first week of February 2020. This is, the relative change in $\Delta KS$ is formulated as $(\Delta KS^{7d}-\Delta KS^{7d}_{Feb.1-7})/\Delta KS^{7d}_{Feb.1-7} \times 100\%$. As can be seen, the three metrics changed after the mandatory lockdown and started to recover to normal levels since the shelter-in-place order was relaxed in June.

Fig.~4B depicts the dynamic evolution of urban structure in each city during the study period. In the initial stage of COVID-19, people kept their commuting as usual and the amenities were open. Since the lockdown began, $\Delta KS$ increased in all cities to varying degrees. It means that all cities became monocentric during lockdown time, particularly Barcelona, Sevilla, and Madrid, which were polycentric before the pandemic. More importantly, their shift to monocentricity implies that trips to destinations outside the CBDs were reduced.
After late June, travel restrictions were lifted and economic activities gradually resumed. The country started getting into a ``new normality'', accompanied by other NPIs. In this context, urban structures recovered to their normal levels in most cities, with Alicante being a notable exception. 
Alicante's polycentricity became stronger than before COVID-19. That might be caused by the migration of second-home owners. In the early stage of the pandemic, second-home owners tend to migrate from crowded cities to low-density areas. Therefore, in Alicante, an increase in the number of second-home owners might have occurred after travel restrictions were lifted~\cite{zougal2020stay}.

In Fig.~4C, when the number of infections gradually raised up at the end of February, there are notable fluctuations in $\overline{Rg}$ for several cities (e.g., Sevilla, Malaga, La Coruna, and Granada). This is mainly caused by the emerging long-distance mobility as a response to future lockdowns. In Fig.~4D, we can see that our shelter-in-place indicator $KS_{HBT}$ is very similar among all the cities.
At the end of phase II, all mobility metrics resumed their normal levels before the lockdown. The resumption of $KS_{HBT}$ is much earlier than that of $\overline{Rg}$. Moreover, as the spatial range of mobility depends on not only on the urban form but also on the spatial scale, the studied cities display varying levels of $\overline{Rg}$ in phase III. In contrast, $KS_{HBT}$ displays a more stable behavior during the entire period, so it might be a more reliable measure to capture the mobility restrictions, agnostic of a city's scale.

Next, we explore the role of urban dynamics and mobility behavior during the spread of COVID-19 across the studied cities. To this end, we estimate the effective reproduction number $R_t$ using the \textit{EpiEstim} R package~\cite{cori2013new}. The results are shown in Fig.~S19. We used $R_t$ in a province as the proxy of that in the metroplex (Note~S6). This approach is feasible as most of the population in a province resides in its metroplex (Table~S2). Generally, $R_t$ started high at the beginning of the COVID-19 outbreak and then decreased below $1.0$ about one week after mandatory lockdowns. Afterward, $R_t$ fluctuated in June, indicating that the second wave was coming. We use gradient boosting machine (GBM) to model $R_t$ with the mobility and population variables described before~\cite{ke2017lightgbm}, and assess each predictor's impact using the SHAP package~\cite{lundberg2020local}.
Figs.~4E-G present each feature's SHAP value distributions for each of the COVID-19 phases defined before. The larger its absolute value is, the stronger a feature impacts $R_t$. The conjunction of high feature value (in red) and positive SHAP value implies that the impact is salient and positive. The mixture of red and blue dots implies an indeterminate impact on $R_t$. The insets in these figures present the relative importance of each factor, measured by the ratio between means of its absolute SHAP values and the average $R_t$ in the phase (see \textit{Assessing feature importance with SHAP} in Materials and Methods). 

As illustrated in Fig.~4E, the mobility variables ($\overline{Rg}$ and $KS_{HBT}$) have stronger and positive impacts on the $R_t$, because phase I covers the early stage of COVID-19 as well as the mandatory lockdown period. In this phase, $\overline{Rg}$ shows the highest impact, followed by $KS_{HBT}$, while the impact of urban dynamics is relatively weak. These observations confirm that in phase I when people travel more frequently and over longer distances, the epidemic spreads faster. Fig.~4F shows the results in the quiet phase II. In conjunction with the insets, we notice that $KS_{HBT}$ still shows a considerable and positive impact on $R_t$ but the impact of $\overline{Rg}$ is ambiguous and weakened. That demonstrates the role of the newly defined mobility metric. Interestingly, we observe that the weight of the urban structure indicator $\Delta KS$ becomes more important in comparison with phase I. In the results of phase III (the second COVID-19 wave) shown in Fig.~4G, all variables became less significant because non-mobility NPIs were the major protection measures.

\section*{Summary and Discussion}

In the course of urbanization, the complex interaction between human mobility, population, jobs, and facilities shapes diverse urban spatial structures. Conventional measures use static sources such as the distribution of employment, facilities, and population to measure the urban structure.  Characterizing cities between mono- to poly-centric organizations~\cite{bertaud2003spatial,pereira2013urban}. However,  measures rely on census data and other data sources that are updated every few years, so they fail to monitor the urban dynamics that can be directly extracted from mobility behaviors.
In particular, during major emergencies such as the COVID-19 pandemic or financial crises, the urban spatial structure can quickly shift, which can significantly impact people's access to resources. Hence, it is important to measure and quantify the dynamics of the urban structure, to enable timely facility renewal and effective economic regulation.

We seek to understand urban dynamics by studying how people interact with their resources in space. With empirical data, we uncover the relationship between people's mobility behavior and the structure of the urban organization. As a result, we devise a new approach to measure the quasi-real-time urban dynamics through the lens of human mobility. A mobility centrality index $\Delta KS$, is introduced to quantify the the heterogeneity of human mobility in a region. A larger value of $\Delta KS$ indicates a more monocentric urban structure, meaning that people are more likely to travel to the city center. In contrast, a lower value of $\Delta KS$ indicates that the urban structure is more polycentric.

After analyzing twenty-one cities via $\Delta KS$ and the Gini index of population, we map these cities into a two-dimensional space. In this space, the mobility dimension is defined by $\Delta KS$, with cities ranging from polycentric to monocentric. In the population dimension defined by the Gini index, cities vary between dispersed and compact. In monocentric cities, like Santa Cruz, Wuhan, and Lisbon, jobs and resources are centralized in the CBD, resulting in an uneven distribution of the characteristic travel distance, $Rg$. There is more heterogeneity in the mobility behavior, resulting in a large $\Delta KS$. In contrast, the resources are distributed more evenly in polycentric cities, such as San Francisco Bay Area and Los Angeles, therefore the distribution of $Rg$ is more homogeneous, and the value of $\Delta KS$ is lower. 

Taking the COVID-19 pandemic in Spain as a case study, we first unveil that the travel restriction on human mobility impacts different income groups in an inequitable manner. The lower income groups took longer trips than the higher income groups at the peak of the infections. This might indicate their lower share of remote workers and less accessibility to online shopping. 
Further, by analyzing mobile phone data during the eight months after the COVID-19 outbreak in eleven cities in Spain, we find that during the first wave in March and April 2020, their urban structures became more monocentric, as trips were mainly going to the CBDs. When the economic activities reopened in June 2020, urban structures returned to their usual forms before COVID-19 and remained stable during the second wave. That can be explained by the lifting of mobility restrictions and a higher dependency on other NPIs such as social distancing and mask wearing. 
In this context, we propose a second dimensionless metric, the extent of home-based travel$KS_{HBT}$, which can capture the extent of sheltering in place. It measures the share of the population with $Rg$ over a certain threshold and ranges between $0$ and $1$, independent of urban scale or population density. We model the effective reproduction number $R_t$ in eleven Spanish cities, illustrating the effects of the proposed mobility metrics over the $R_t$.

It's noteworthy that, we used the radius of gyration of individual traces in the calculation of the proposed mobility metrics. However, to cope with people's privacy concerns, most smartphone platforms are moving to anonymized aggregate data collection (k-anonymity, etc.) In this context, although $Rg$ only presents the mobility scale of one user, without any private information, we needed the individual trajectory of each user to calculate their $Rg$. Therefore, it is not possible to directly calculate $Rg$ (see Fig.~2A) with K-anonymity. To solve this issue, a specially-designed K-anonymity strategy may be required, or the mobile operator could collect $Rg$ from the smartphone, rather than people's mobility traces.

This work offers several promising avenues for further investigation. Although we have observed that low-income individuals traveled longer distances than their high-income counterparts during the initial COVID-19 outbreak, the purposes of these trips remain unclear. This knowledge gap limits our understanding of the unequal impact of lockdown measures. To address this, future research could leverage long-term mobility traces to identify essential workers and explore their behavior during emergency events, with a focus on socio-economic equity.

Taken together, our work highlights the value of fine-grained individual mobility data for quantifying urban dynamics. The near real-time  sensing of urban dynamics becomes more important in large-scale emergency events, like pandemics or natural disasters, to better plan for the reopening or reconstruction of cities.

\section*{\LARGE{Materials and Methods}}
\subsection*{Datasets description}

\begin{enumerate}
\item[] {\bf Demographic data} The population with a spatial resolution of 30 arc-seconds (approximately 1 km$^2$ near the equator) of each city was obtained from the LandScan~\cite{landscan2015} in 2015.
To compare the residents of different socio-demographic groups for the United States urban areas, we classify tracts into poor and non-poor based on criteria that a household's median income is less than the national middle-class threshold $\$45,200$. Similarly, we classify tracts as majority non-Hispanic white, non-Hispanic black, Hispanic, or Asian using a threshold of $50\%$. Racial and economic data are from the 2015 American Community Survey (ACS 2015) aggregated at census tract level~\cite{Census}. Since in Boston there are too few Asian tracts to permit reliable analyses for that group we decided to not include it in this analysis. Regarding Bogot\'a, we employ the socio-economic strata (SES) classification used by the city administration. Every tract is classified into a SES score, from 1 to 6, representing resident's income level from low to high, being six being the highest. 

\item[] {\bf Mobility data} Among the twenty-one cities, we leverage the travel survey data including the home locations and daily visited locations, recorded among 567,301 and 444,127 users in Atlanta and Rio de Janeiro, respectively. In Boston, SFBay, LA, Bogot\'a, Lisbon and Porto, we apply the TimeGeo framework to model individuals' travel behavior at a fine granularity, e.g., every 10 minutes, with the CDRs data~\cite{jiang2016timegeo}. TimeGeo extracts stay points from each individual’s sequence of records, and separates commuters from non-commuters by checking users' work locations. For each census tract (or equivalent unit area) in the metropolitan area expansion factors are calculated for commuters and non-commuters using census data. Based on the distributions of the empirical mobility parameters extracted from the active user data, a simulation of how the entire urban population move is achieved. For some of the cities, the results have been validated with some travel survey data~\cite{jiang2016timegeo,xu2018planning,de2020socio}. 
For other cities, including Wuhan, Shenzhen and all the Spanish cities, we leverage the XDRs data to extract the visited locations of the active users. All data are anonymized by mobile phone operators for privacy protection and each record provides the anonymized user id, the time and geographical location of using cell phone. As the home locations of users are not available in either CDRs or XDRs, we identify users' most frequently visited location during weekends and weekdays nights as their home locations.

\end{enumerate}

\subsection*{Estimate of $R_t$ from reported cases}
The \textit{EpiEstim} R package was developed by Cori and colleagues~\cite{cori2013new}, and has been adopted to estimate the transmission intensity of SARS-CoV-2 in varying countries~\cite{abbott2020estimating,nouvellet2021reduction}. \textit{EpiEstim} provides a way to measure the reproduction number $R_t$ of an epidemic based on the daily number of new infections. Due to the delayed reports and limit accuracy of epidemiological data, there are several hypothesises in the solution. First, \textit{EpiEstim} uses a Bayesian inference framework to estimate the posterior probability of $R_t$, assuming its prior probability as a Gamma-distribution. Based on the results reported by Imai et al.~\cite{imai2020report}, we used a prior Gamma distribution for $R_{t,\tau}$ with mean of $2.6$ and a standard deviation of $2.0$. Second, \textit{EpiEstim} requires the distribution of \textit{serial interval}, which means the time difference between the onset of the symptoms of a primary case and her corresponding secondary cases. This work follows the observation from~\cite{ryu2022serial}, and assume \textit{serial interval} as a discrete Gamma-distribution, with mean of $3.6$ days and a standard deviation of $4.9$ days. Third, in order to achieve stable estimation, we used a seven-day ($\tau=7$ days) time window before the day $t$ to calculate $R_t$ for each province in Spain. The estimated $R_t$ and the $95\%$ confidence interval in each city are illustrated in Fig.~S19. More details about the implementation of \textit{EpiEstim} can be found in Note~S6 and Ref.~\cite{cori2013new}.

\subsection*{Assessing feature importance with SHAP}
For the selected Spanish cities, we adopt the SHAP package to assess the impacts of the mobility and urban dynamics variables to the effective reproduction number $R_t$, which serves as the proxy for how COVID-19 spreads. SHAP is a game theoretic approach to explain the output of a given machine learning model~\cite{lundberg2020local}. To build a regression model for $R_t$, we utilize the gradient boosting machine (GBM) implemented by the LightGBM package~\cite{ke2017lightgbm}. More specifically, we collect the time-varying variables ($\overline{Rg}^{7d}$, $KS_{HBT}^{7d}$, $\Delta KS^{7d}$) and the constant variables ($\overline{Rg}_{typ}$, $\Delta KS_{typ}$, logarithmic value of the total population, and Gini value of population distribution) in the cities, serve them as the input to the GBM model, and set the corresponding $R_t$ as the output. With the GBM model, the $R_t$ can be well modeled in each phase of COVID-19 with high $r^2$ values ($r^2=0.98$, $0.94$, and $0.85$ for the three phases). 

Given this well-trained GBM model, we can use the SHAP package to quantify how the factors in GBM model contribute to the prediction of $R_t$ for each city on a daily basis. The quantification is done with a namely SHAP value, as shown in Figs.~4E-G. A large absolute SHAP value implies that the prediction of $R_t$ would be more sensitive to the given factor in the GBM regression model. In these charts, the x-axis stands for SHAP values of the features in the modeling of $R_t$, and the y-axis has all the mobility and urban variables we input into GBM. Each colored point on the chart indicates the SHAP value of the feature in the prediction of $R_t$. Red and blue colors imply higher and lower values of a feature respectively. Thus, we can read features' directionality impact to $R_t$ in the well-trained GBM model based on the distribution of the red and blue dots. Taking the SHAP values of $\overline{Rg}^{7d}$ in Fig.~4E as an example, we can see higher value of $\overline{Rg}^{7d}$ leads to higher $R_t$ in GBM model, and lower value of $\overline{Rg}^{7d}$ leads to lower value of $R_t$. That is, $\overline{Rg}^{7d}$ shows clearly positive relation to the spread of COVID-19 in phase I.
Next, for each phase of COVID-19, we aggregate the SHAP values of all samples to assess the importance of each factor holistically, as shown in the insets of Figs.~4E-G. To achieve the relative importance of each factor, we calculate the mean of the absolute SHAP values and then divide it by the average $R_t$ for that phase.


\bibliography{scibib}
\bibliographystyle{Science}

\section*{Acknowledgments}
We thank Prof. Pu Wang for the data provided. This work was supported by the Berkeley DeepDrive (BDD) and the ITS-SB1 Berkeley Statewide Transportation Research Program. Yanyan Xu and Xiaokang Yang were also supported by the National Natural Science Foundation of China (62102258), the Shanghai Pujiang Program (21PJ1407300), the Shanghai Municipal Science and Technology Major Project (2021SHZDZX0102), and the Fundamental Research Funds for the Central Universities.

\section*{Author contributions}
YX, LEO, XY and MCG conceived the research and designed the analyses. DM and AH processed the Spanish data. YX and LEO performed the analyses. MCG and YX wrote the paper. MCG supervised the research.

\section*{Inclusion \& ethics}
All contributors fulfill the authorship criteria are listed as co-authors in this paper. Other contributors who do not meet all criteria for authorship are listed in the Acknowledgements.

\section*{Competing interests}
The authors declare that they have no competing interests. 

\section*{Data and materials availability} 
All data needed to evaluate the conclusions in the paper are described in the paper and the Supplementary Materials. For contractual and privacy reasons, we cannot make the raw mobile phone data available. A sample of the data may be provided by the corresponding author upon reasonable request. 

\section*{Code availability}
The implementation of this work and sample data are available at:\\ \url{https://github.com/humnetlab/UrbanForm}.

 
\section*{Supplementary materials}
Notes S1 to S6 \\
Figs. S1 to S19 \\
Table S1 and S2 \\


\clearpage

\begin{figure*}[htb!]
\centering
\includegraphics[width=0.9\linewidth]{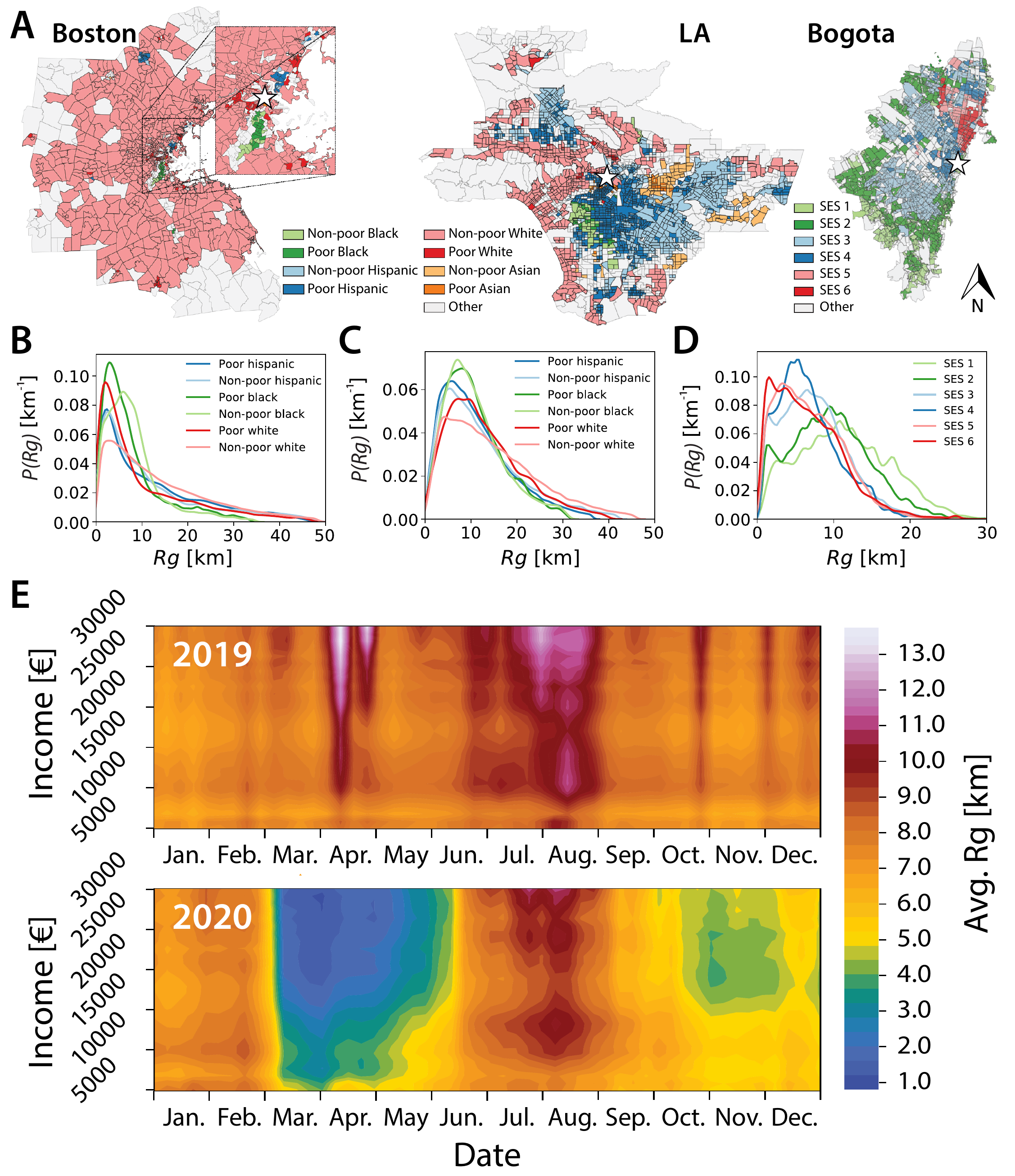}
\end{figure*}
\noindent{{\bf Fig. 1. Socio-economic segregation and its relation to radius of gyration.} (\textbf{A}) Spatial segregation of residents' socio-economic statuses in Boston, LA, and Bogot\'a. White star in each map indicates the location of CBD. High-income residents generally live in the suburbs (light colors) in Boston and LA, resulting in low-income areas concentrated in the central city (dark colors). This pattern is reversed in Bogot\'a, where higher-income groups (SES 5, 6) live close to the city’s CBD and lower-income ones tend to live in the periphery. (\textbf{B-D}) Distribution of $Rg$ for different socio-economic groups in Boston, LA, and Bogot\'a. High-income residents in Boston and LA tend to travel farther from their home compared to the low-income residents. The inverse phenomenon is observed in Bogot\'a. (\textbf{E}) Relation between annual income and average $Rg$ of the population in Spain during 2019 and 2020, covering the periods before and during the COVID-19. High-income residents take more long-distance trips during the tourist seasons, e.g., April and summer of 2019, summer of 2020. In contrast, low-income residents have longer travel distances during the peak periods of COVID-19, e.g., spring and autumn of 2020.}

\newpage

\begin{figure*}[tb!]
\centering
\includegraphics[width=1.0\linewidth]{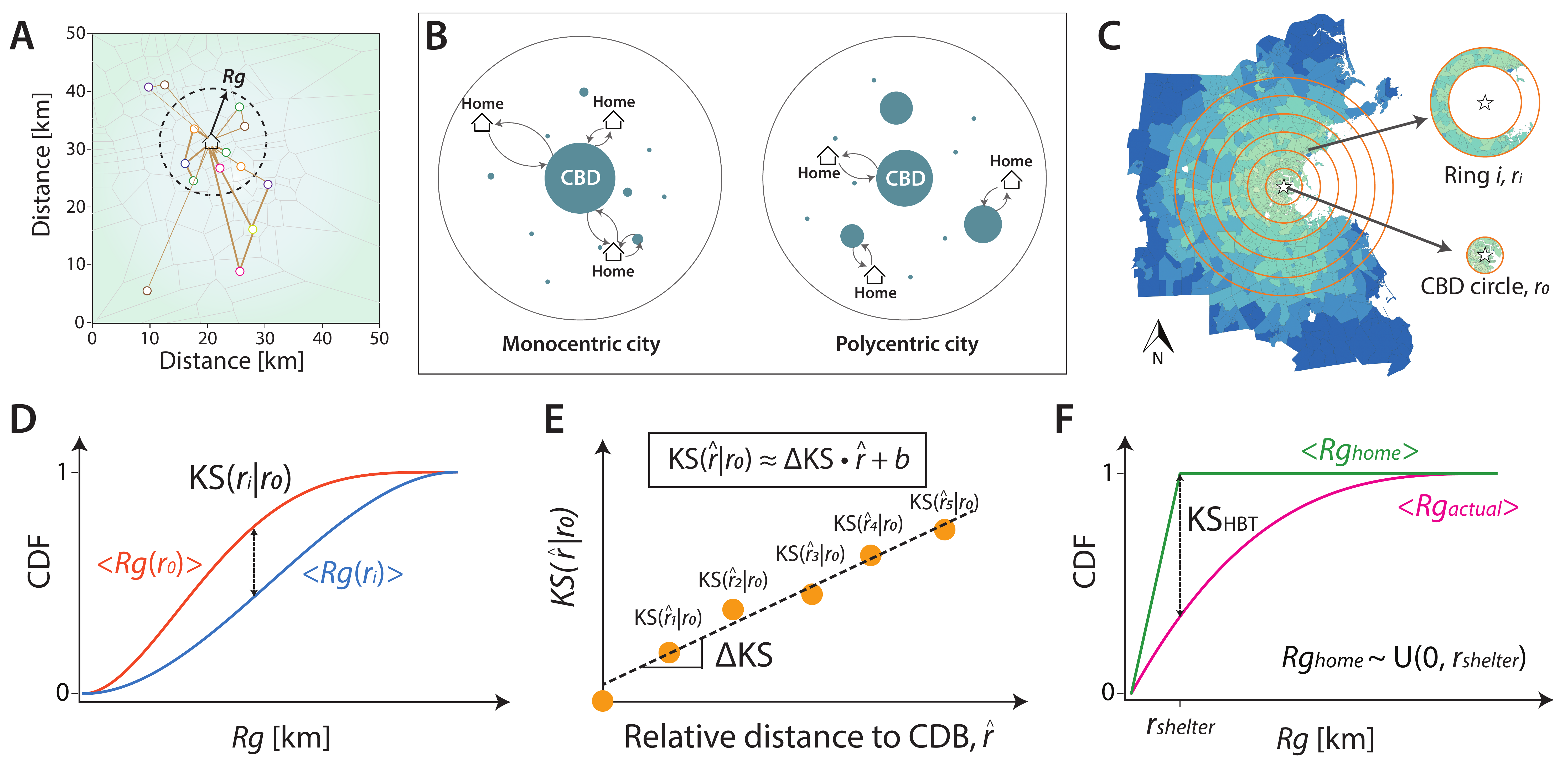}
\end{figure*}
\noindent{{\bf Fig. 2. Schematic illustration of aggregate metrics relating human mobility to urban structure.} (\textbf{A}) Radius of gyration, $Rg$, representing the spatial coverage of the locations visited by an individual during the period considered. (\textbf{B}) Illustration of typical mobility behavior in monocentric and polycentric cities. People in monocentric cities tend to travel to the CBD to acquire job or other resources, while in polycentric cities, the CBDs are less attractive and people's travel destinations are distributed more dispersedly in the space. (\textbf{C}) Illustration of the choropleth map in Boston. The concentric rings have incremental radii. The width of each ring $\Delta r$ is set as $3$~km for all of the cities. (\textbf{D}) Definition of $KS(r_i|r_0)$, the Kolmogorov–Smirnov statistic, quantifying the distribution disparity of the $Rg$ in the peripheral ring $r_i$ and the CBD circle $r_0$. (\textbf{E}) Definition of mobility centrality index $\Delta KS$, representing the spatial variation of residents' $Rg$ with respect to their distances to the CBD. (\textbf{F}) Definition of the home-based-travel indicator $KS_{HBT}$, referring to the Kolmogorov–Smirnov statistic between the actual $Rg$ of population and their $Rg$ under travel restriction.}

\newpage

\begin{figure*}[tb!]
\centering
\includegraphics[width=1.0\linewidth]{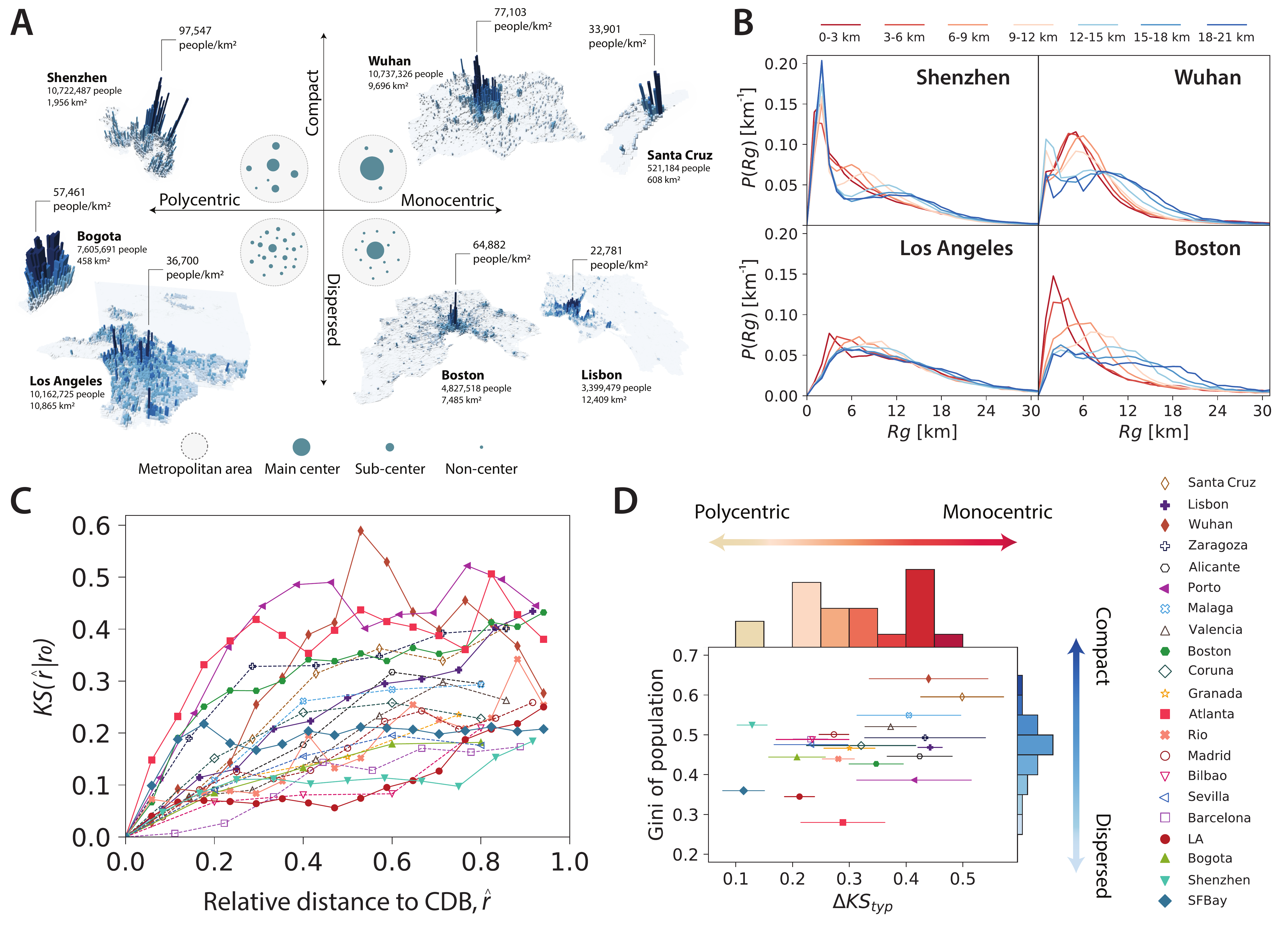}
\end{figure*}
\noindent{{\bf Fig. 3. Defining urban structure via mobility behavior of the population.} (\textbf{A}) Two dimensions of urban spatial structure. Centrality, ranging from polycentricity to monocentricity, reflects the number of population centers. Concentration, ranging from dispersion to compaction, refers to the degree of heterogeneity or dispersion in the amount of population shared outside the CBD (adapted from~\cite{meijers2010spatial}). The population distribution of some sample cities are visualized with 3D barplots. (\textbf{B}) Distribution of $Rg$ values in seven rings in Shenzhen, Wuhan, LA, and Boston. In Shenzhen and LA, the population residing in varying distances to the CBD exhibit similar distributions of $Rg$; whereas in Wuhan and Boston, the population residing in outer rings tends to have longer travel distances from their home. (\textbf{C}) Relation between mobility distribution changes $\Delta KS(\hat{r}|r_0)$ in peripheral rings and their relative radii to the CBD. All cities exhibit positive relations. (\textbf{D}) Distribution of cities in the two-dimension space defined by $\Delta KS_{typ}$ and Gini index of population. The error bar of each city shows the standard error of linear fitting for $\Delta KS$ calculation. Cities in the legend are shown in descending order of $\Delta KS$. Among the twenty-one cities, Wuhan, Santa Cruz and Malaga are clearly compact and monocentric; San Francisco Bay Area and Los Angeles are clearly dispersed and polycentric; Shenzhen is compact but polycentric.}

\newpage

\begin{figure*}[htb!]
\centering
\includegraphics[width=0.95\linewidth]{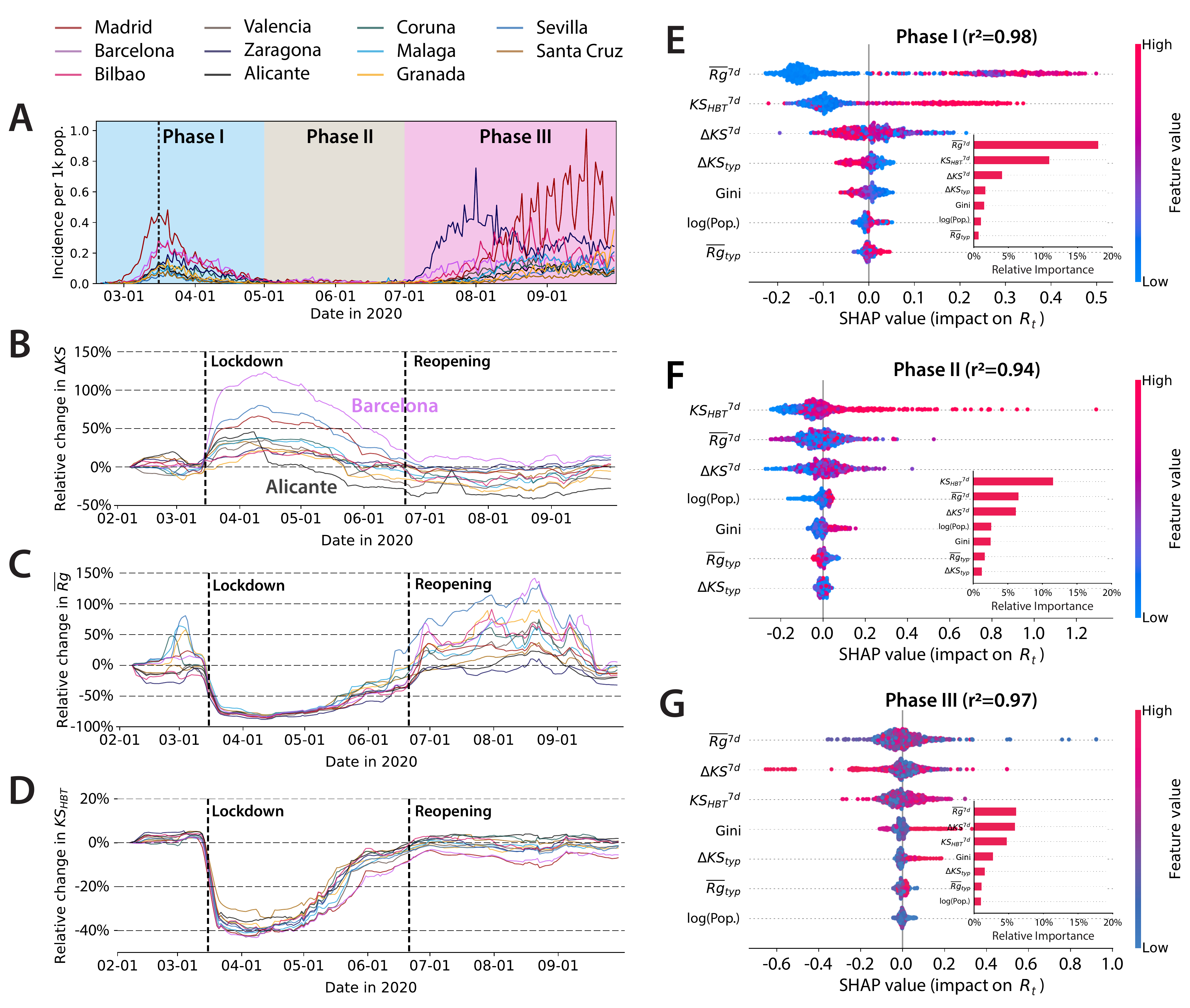}
\end{figure*}
\noindent{{\bf Fig. 4. Changes in urban form and mobility behavior during the COVID-19 pandemic in the eleven Spanish cities.} (\textbf{A}) Daily reported infections per 1,000 population in each province by September 30, 2020. Nationwide lockdown was imposed on March 14th, marked with the dashed line. We divide the entire period into three phases: the first wave, a quiet period, and the second wave. (\textbf{B}) Relative change of mobility centrality index $\Delta KS$, with respect to the first week in February. Almost all cities became more monocentric during the phase I and recovered to normal level during the phases II and III. (\textbf{C}) Relative change of mobility scale measure by $\overline{Rg}$. (\textbf{D}) Relative change of the home-based-travel indicator $KS_{HBT}$. Both of the $\overline{Rg}$ and $KS_{HBT}$ sharply decreased after the mandatory lockdown was imposed at national scale. The $KS_{HBT}$ values recovered to normal level after the lockdown was lifted, while $\overline{Rg}$ fluctuated irregularly in these cities. (\textbf{E-G}) Impacts of the mobility behavior on the effective reproduction number $R_t$ during the three phases. The scatter plots display the impact of each factor on the $R_t$ on a given day and in one city. The insets illustrate the relative importance of each factor, represented by the ratio between means of their absolute SHAP values and the average $R_t$ during each phase.}

\end{document}